\documentclass{aa}  
 \usepackage{comment} 
\usepackage{colortbl}
\usepackage{graphicx}
\usepackage{txfonts}
\usepackage[]{caption}
\usepackage{subfigure}
\usepackage{multirow}
\definecolor{pink}{rgb}{1.0, 0, 0.8}
\usepackage{tabularx}
   
\begin{document} 
\title{All about Cookies:\\The perfect compromise between softness and crispiness}
\author{S.\ Rosu}
\mail{sophie.rosu@unige.ch}
\institute{Department of Astronomy, University of Geneva, Chemin Pegasi 51, 1290 Versoix, Switzerland\\sophie.rosu@unige.ch}
\mail{sophie.rosu@unige.ch}
\date{April 1, 2025}

\abstract{Cookies are enjoyed best when they are both crispy and soft.}{I investigate in which proportion the cookies are crispy and soft, and disentangle whether it makes them biscuits, cakes, or none of the above.}{I baked cookies for colleagues at KTH, Stockholm, and University of Geneva, Switzerland, adopting my mum's mum's mum's etc. recipe. I created a dedicated survey for my colleagues with three well-selected questions to answer while eating one cookie.}{The weighted-average mean of the crispiness and softness, weighted by the respective enjoyment of the cookie, over the whole population amount to $7.0\pm1.1$ and $5.3\pm1.4$, respectively. The enjoyment of the cookies amounts to $9.1\pm2.3$.}{People like (my) cookies, and cookies are neither cakes, nor biscuits, they are just... cookies!} 

\maketitle

\section{Introduction and motivation of this work}
Recently (i.e. 4 years ago), I read a mind-blowing paper about the Jaffa Cake \citep{stevance21}. The author put the debate about their classification as a biscuit (based on their size and host environment) or as a cake (based on their physical properties) to rest. While it is pretty clear that biscuits are 100\% crispy, cakes can be 100\% soft, 100\% crispy, or be somewhere in between, depending how much time you have forgotten them on your shelf. However, I, the reader, was left at that time hungry for more: What about cookies? I could not find any such thorough study about cookies; this is why I undertook it (after a few years, but I had other \textit{less important} business about binary stars to finish).

It is rather common knowledge that people want cookies to be both crispy and soft. But in which proportion? Does that make them biscuits or cakes? To answer these questions, I have built a dedicated experimental set up (Sect.\,\ref{sect:methodology}): I baked cookies (Sect.\,\ref{subsect:recipe}) and brought them to my colleagues whom I used as guinea pigs to answer my survey (Sect.\,\ref{subsect:survey}). The results are presented in Sect.\,\ref{sect:results} and discussed in Sect.\,\ref{sect:discussion}.

\section{Methodology}\label{sect:methodology}
\subsection{Cookies preparation}\label{subsect:recipe}
The cookies' recipe I follow is my mum's recipe, which is itself her mum's recipe, itself her mum's mum's recipe, etc. In other words, if we go as far as we can (i.e. when basic ingredients such as chocolat were produced), we all have the same recipe -- which we know is not true. Therefore, here is what I will call for now on, \textit{my} recipe of cookies.\\

\begin{figure}[h]
\includegraphics[clip=true, trim=200 200 250 380,width=\linewidth]{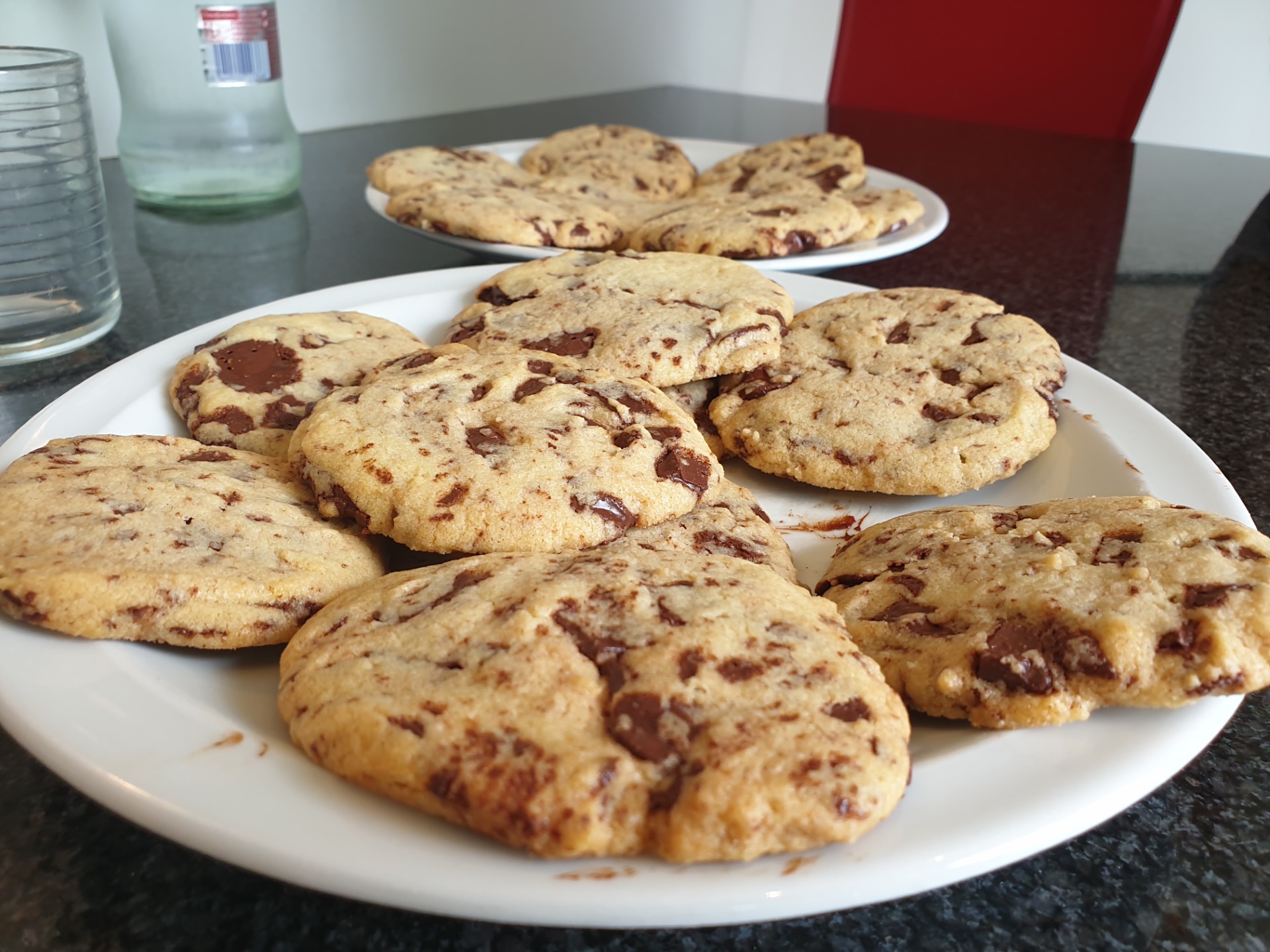}
\caption{Cookies baked in a 40 years old oven following the recipe presented in Sect.\,\ref{subsect:recipe}. \copyright Me.\label{fig:cookie}}
\end{figure}

\noindent Ingredients: 
\begin{itemize}
\item[$\bullet$] 230 g of flour
\item[$\bullet$] 90 g of sugar
\item[$\bullet$] 90 g of brown sugar
\item[$\bullet$] 10 g of vanilla sugar
\item[$\bullet$] 1 teaspoon of baking powder
\item[$\bullet$] 1 egg
\item[$\bullet$] 1 pinch of salt
\item[$\bullet$] 125 g of butter
\item[$\bullet$] 200 g of dark (54\%) chocolate
\end{itemize} 
Preparation:
\begin{enumerate}
\item Preheat the oven at 140-150$^\circ$C; 
\item Soften the butter in the microwave or in the oven; 
\item Mix the soften butter with the sugar, brown sugar, vanilla sugar, baking powder, salt, egg, and flour with a hand mixer;
\item Break the chocolate in small pieces of about 1 cm$^3$, add it to the previous preparation, and mix with the hands\footnote{Previously washed (the hands, I mean), unless you want to add some more taste in the preparation.}; 
\item Place a baking paper on a plate that goes into the oven; 
\item Shape small clods of dough of approximately one tablespoon and place them on the baking paper, separated by a distance of at least their size\footnote{It is like binary stars: give them large-separation orbits, otherwise they will soon fill their Roche lobe and merge, and you will get only one giant cookie.};
\item Bake the cookies for 25 minutes (but see Sect.\,\ref{sect:discussion});
\item Eat them warm or let them cool down.
\end{enumerate}

The results are depicted in Fig.\,\ref{fig:cookie}.

\subsection{Survey}\label{subsect:survey}
The survey was not confidential but there has been no phishing afterwards. The participants had to accept cookies before proceeding, even if they did not want to. While eating their -- I hope respective -- cookie, the participants were asked to answer three questions on a Post-it note someone in the audience collected afterwards in a roughly anonymous manner. The participants were asked to give honest answers (even for the last question) while the author of this paper looked at them intensively, like surveying a test. The three questions are listed below with their respective scales to facilitate the participants' choice:
\begin{enumerate}
\item How crispy is the cookie on its exterior?
\begin{itemize} 
\item[$\bullet$] 0: As much as a soft caramel;
\item[$\bullet$] 1-9: Left to the participant's appreciation\footnote{It does absolutely not remind me any survey I have answered the last years.};
\item[$\bullet$] 10: Even the dust on my shelves is softer;
\end{itemize}
\item How soft is the cookie in its interior?
\begin{itemize} 
\item[$\bullet$] 0: I broke my tooth, I need to go to the dentist now;
\item[$\bullet$] 1-9: Left to the participant's appreciation;
\item[$\bullet$]10: Even water is less liquid;
\end{itemize}
\item How much do you enjoy it?
\begin{itemize}
\item[$\bullet$] 0: I will soon pretend I have a meeting and put this one in the trash;
\item[$\bullet$] 1-9: Left to the participant's appreciation;
\item[$\bullet$] 10: Polite or not, I do not care, I will take the last one and all others before.
\end{itemize}
\end{enumerate}

The survey was carried out in June 2024 in the Particle, Astrophys \& Med Imag department at KTH Royal Institute of Technology, Stockholm, Sweden (called `KTH' for now on) with 19 participants, and in March 2025 in the Stellar group in the Department of Astronomy in Geneva, Switzerland (called `UNIGE' for now on) with 19 participants (me included).

\section{Results}\label{sect:results}
\begin{figure}[h!]
\includegraphics[clip=true, trim=10 0 30 30,width=\linewidth]{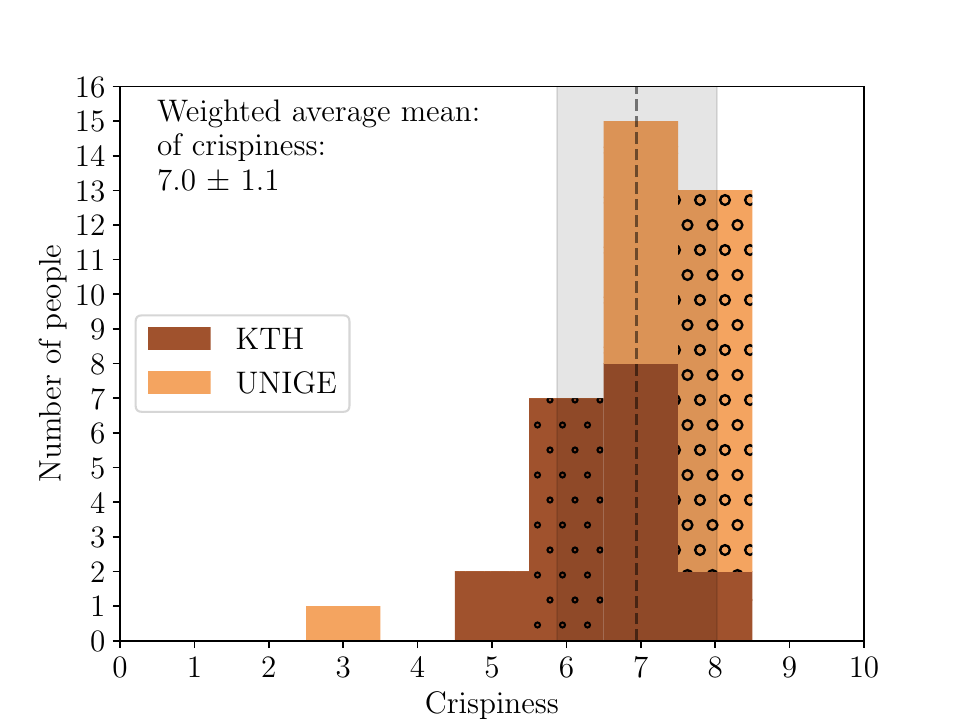}
\includegraphics[clip=true, trim=10 0 30 30,width=\linewidth]{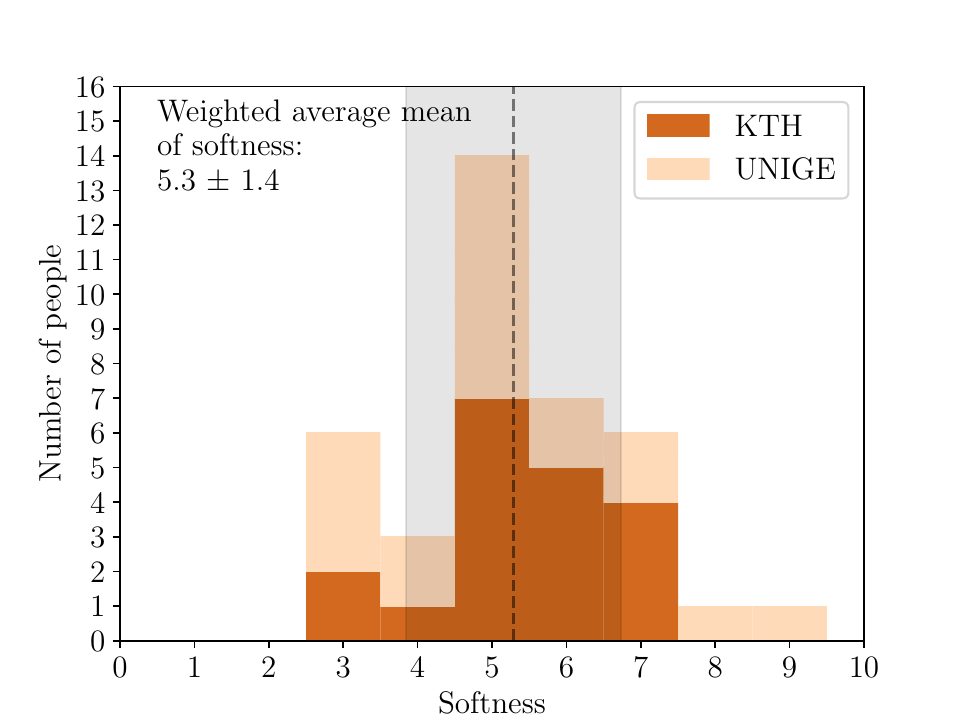}
\includegraphics[clip=true, trim=10 0 30 30,width=\linewidth]{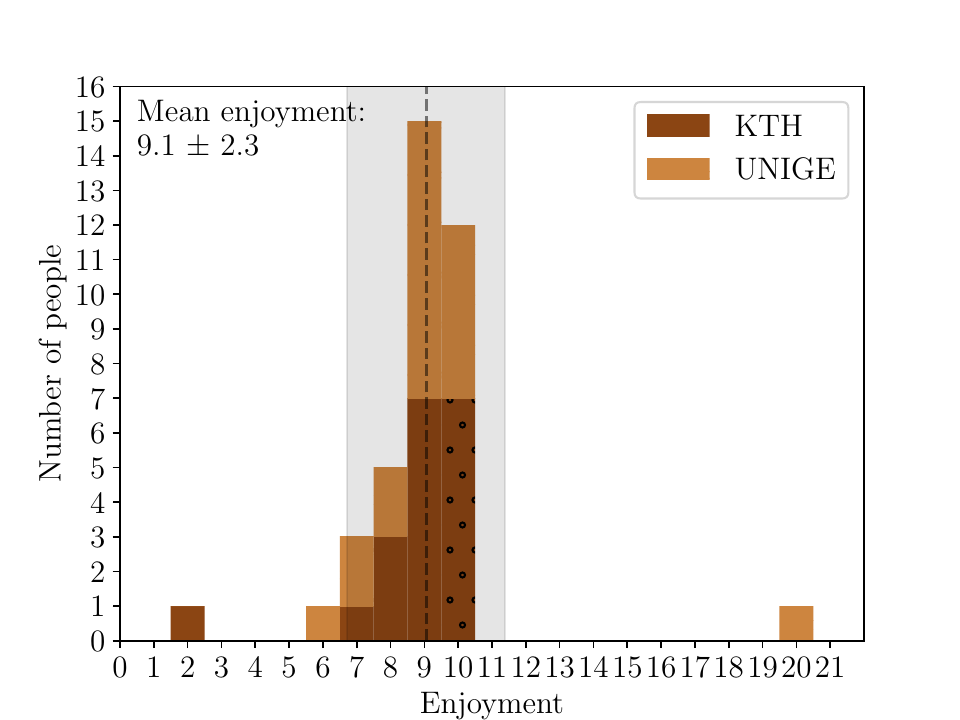}
\caption{Histograms of the crispiness (top panel), softness (middle panel), and enjoyment (bottom panel). The values indicated in the plots are global values over KTH and UNIGE. \label{fig:histograms}}
\end{figure}

\begin{figure}[t]
\includegraphics[clip=true, trim=0 0 0 0,width=\linewidth]{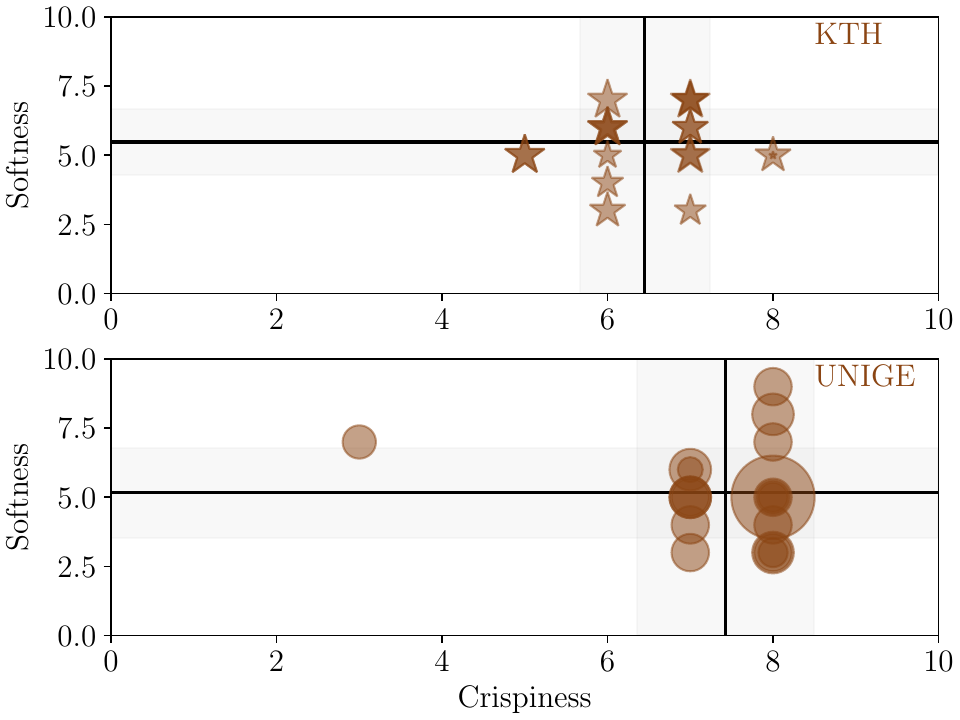}
\caption{Softness as a function of crispiness of the cookies for KTH (top panel) and UNIGE (bottom panel). Each star/circle represents one cookie; its size is a direct function of the associated enjoyment of the cookie.\label{fig:softness_crispiness}}
\end{figure}

The distribution of results in terms of crispiness, softness, and enjoyment of the cookies are presented in Fig.\,\ref{fig:histograms}, and separated by survey (KTH and UNIGE). To estimate the average and standard deviation of the crispiness and softness, I decided to take the weighted-average mean of these two quantities, weighted by the respective enjoyment of the cookie. Indeed, the purpose here is not to determine how crispy and soft my cookies are, but rather how crispy and soft people want the cookies to be in general. In such a way, the higher the enjoyment, the higher the weight\footnote{To compute the mean. Not the weight of the person. I hope.}! A simple mean and standard deviation were computed for the enjoyment. The results are summarised in Table\,\ref{table:values}. In Fig.\,\ref{fig:softness_crispiness}, I present the cookies' softness as a function of crispiness for KTH and UNIGE. 

\begin{table}[t]
\centering
\begin{tabular}{lllll}
\hline \hline
& KTH & UNIGE & Global \\
\hline 
Crispiness & $6.5\pm 0.8$ & $7.4\pm 1.1$& $7.0 \pm 1.1$ \\
Softness & $5.5 \pm 1.2$ & $5.2 \pm 1.6$ & $5.3 \pm 1.4$ \\
Enjoyment & $8.7. \pm 1.8$ & $9.4 \pm 2.7$ & $9.1\pm2.3$ \\
\hline 
\end{tabular}
\caption{Values of the crispiness, softness, and enjoyment obtained at KTH, UNIGE, and globally. The crispiness and softness averages are weighted by the enjoyment (see Sect.\,\ref{sect:results} for details). \label{table:values}}
\end{table}

Globally, the best cookies should have a crispiness of $7.0\pm 1.1$, a softness of $5.3\pm1.4$, and an enjoyment of $9.1\pm2.3$. The best crispiness and softness, as well as the enjoyment of the cookies between KTH and UNIGE agree very well within the error bars. Though, we note a general trend of cookies at UNIGE being more crispy both in their exterior and interior. Yet, it does not make the cookies less good, given the higher enjoyment of people at UNIGE than at KTH. Still, the higher enjoyment at UNIGE is driven by the outlier (definitely a very kind person who decided to mess up my survey by answering outside of the box, there is always that one person). While the crispiness and the softness of the cookies are more evenly distributed around their respective means at KTH, there is a clear convergence towards a value of 7-8 for the crispiness of the cookies at UNIGE, with only one outlier giving a value of 3.

\begin{figure}[t]
\includegraphics[clip=true, trim=0 0 0 0,width=\linewidth]{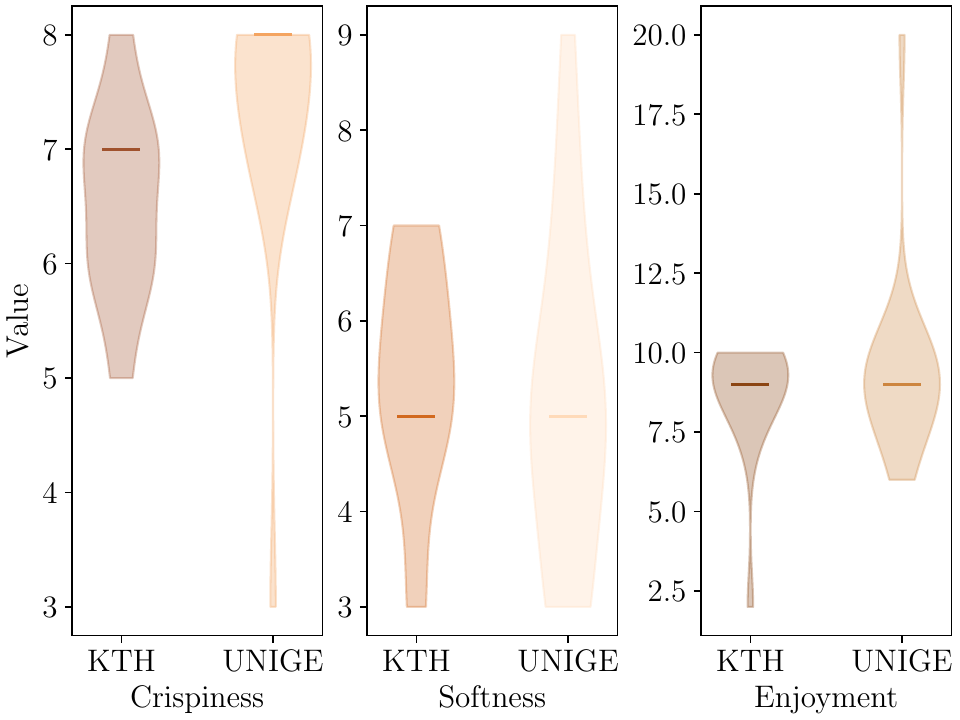}
\caption{Violin plot series of the crispiness (left panel), softness (middle panel), and enjoyment (right panel) for KTH and UNIGE. \label{fig:violin}}
\end{figure}

A probably better way to visualise the results is shown in Fig\,\ref{fig:violin}: this figure is so clear that I do not even dare to analyse it.

\section{Discussion and conclusion}\label{sect:discussion}
The contrast between a crispy exterior and a soft interior makes the cookie extremely difficult to put in either of the biscuit or cake categories. For this reason, I invite the community to simply call them cookies. 

Another very personal conclusion is that my cookies are definitely good, with an enjoyment of $9.1\pm2.3$ (as the survey was not at all biased). More precisely, an exterior crispiness of $7.0\pm 1.1$ and an interior softness of $5.3\pm 1.4$ seem to be the perfect combination for a cookie. The higher crispiness observed for both the exterior and interior of the cookies at UNIGE is probably attributable to the different oven I used in Switzerland: I had to severely decrease the temperature of the oven (140$^\circ$C instead of 150$^\circ$C) and the cooking time (20 minutes instead of 25 minutes), and the cookies were still more baked than in Sweden (as tested by myself). 

As a future work, I plan to bake more cookies to improve the statistics (though I already have more than two points, so my conclusions should be robust enough). If you would like to be a volunteer, please contact me and I will send you the cookie by post: It will enable me to extend the survey to mouldy cookies, delicious! Finally, I will make a follow-up to find this \&\#\$\%@ person who dared to put a 2/10 for the enjoyment of my cookies!

 \begin{acknowledgements}
I would like to thank my colleagues at the Particle, Astrophys \& Med Imag department at KTH Royal Institute of Technology, Stockholm, Sweden for challenging me coming up with a science case about cookies in approximately three hours to be presented at our regular Friday's Fika (if you do not know what the Fika concept is, visit this \url{https://visitsweden.com/what-to-do/food-drink/swedish-kitchen/all-about-swedish-fika/}). For the sake of their career, I will not name them. I also warmly thank both my colleagues at KTH and in the Stellar group in the Department of Astronomy in Geneva, Switzerland to have dedicated half an hour of their time and accepted to take part in this survey. I would like to thank my colleague (who I will also not name for his sake) for providing suggestions to improve this manuscript. Finally, I would like to thank in advance my future employer(s) to \textbf{not} take into account this paper in my list of publications. I hope you, as a reader, had as much pleasure reading this paper as I had writing it.\\

\textit{Ethics related.} I here declare there was no human abuse (there was abuse of cookies by few people, but that is another story) and that this survey is ethically correct: people were given the same quantity of cookies, within $1\sigma = 1 g$. 

\end{acknowledgements}

\end{document}